\documentclass[shortnote,twocolumn]{jpsj3-arxiv}

\usepackage{txfonts}
\usepackage[dvipdfmx]{graphicx,color}
\usepackage{dcolumn}
\usepackage{bm}
\usepackage{braket}
\usepackage[version=3]{mhchem}
\usepackage{url}
\usepackage{cite}
\usepackage{ulem,color}
\usepackage{mathrsfs}

\newcommand{\mrm}{\mathrm}

\newcommand{\mcl}{\mathcal}

\newcommand{\tht}{\theta}

\newcommand{\ham}{\mcl{H}}

\newcommand{\pri}{\prime}

\newcommand{\figref}[1]{Fig.~\ref{#1}}

\title{Revisit of magnetic orders in 1/1 approximant crystals of Tsai-type quasicrystal from theoretical points of view}

\author{Takanori Sugimoto$^1$\thanks{sugimoto.takanori.qiqb@osaka-u.ac.jp}, Shintaro Suzuki$^2$, Ryuji Tamura$^3$, Takami Tohyama$^4$}
\inst{$^1$Center for Quantum Information and Quantum Biology, Osaka University, Toyonaka, Osaka 560-0043, Japan \\
$^2$Department of Physical Science, Aoyama Gakuin University, Sagamihara, Kanagawa 252-5258, Japan \\
$^3$Department of Materials Science and Technology, Tokyo University of Science, Katsushika, Tokyo 125-8585, Japan \\
$^4$Department of Applied Physics, Tokyo University of Science, Katsushika, Tokyo 125-8585, Japan
} 

\abst{The various magnetic orders observed in approximant crystals of Tsai-type quasicrystal are important for deepening our understanding of mysterious magnetism in quasicrystals.
Using an icosahedral cluster model with inter-cluster interactions, we give an intuitive explanation of the mechanism of magnetic orders previously reported by classical Monte-Carlo simulations for an approximant-crystal model describing a Gd-based Tsai-type 1/1 approximant crystal.}

\begin{document}
\maketitle

Recent experimental discovery of ferromagnetism in quasicrystals (QCs) has stimulated both experimental and theoretical investigations of peculiar magnetism inherent in QCs~\cite{Tamura2021,Takeuchi2023}.
However, microscopic magnetic structure and mechanism of the ferromagnetic order have not been clarified so far.
For solving this problem, {\it approximant crystals} (ACs) of the QCs can provide crucial information as emphasized before~\cite{Goldman2014,Suzuki2021-2,Sato2022}.
In fact, recent comprehensive experimental studies on the magnetism of  ACs~\cite{Suzuki2021-2,Hiroto2013,Hiroto2014,Ishikawa2016,Ishikawa2018,Yoshida2019,Labib2020,Inagaki2020,Labib2022} are helpful for understanding the mechanism of ferromagnetism in QCs.

\begin{figure}[htb]
\centering
\includegraphics[width=0.50\textwidth]{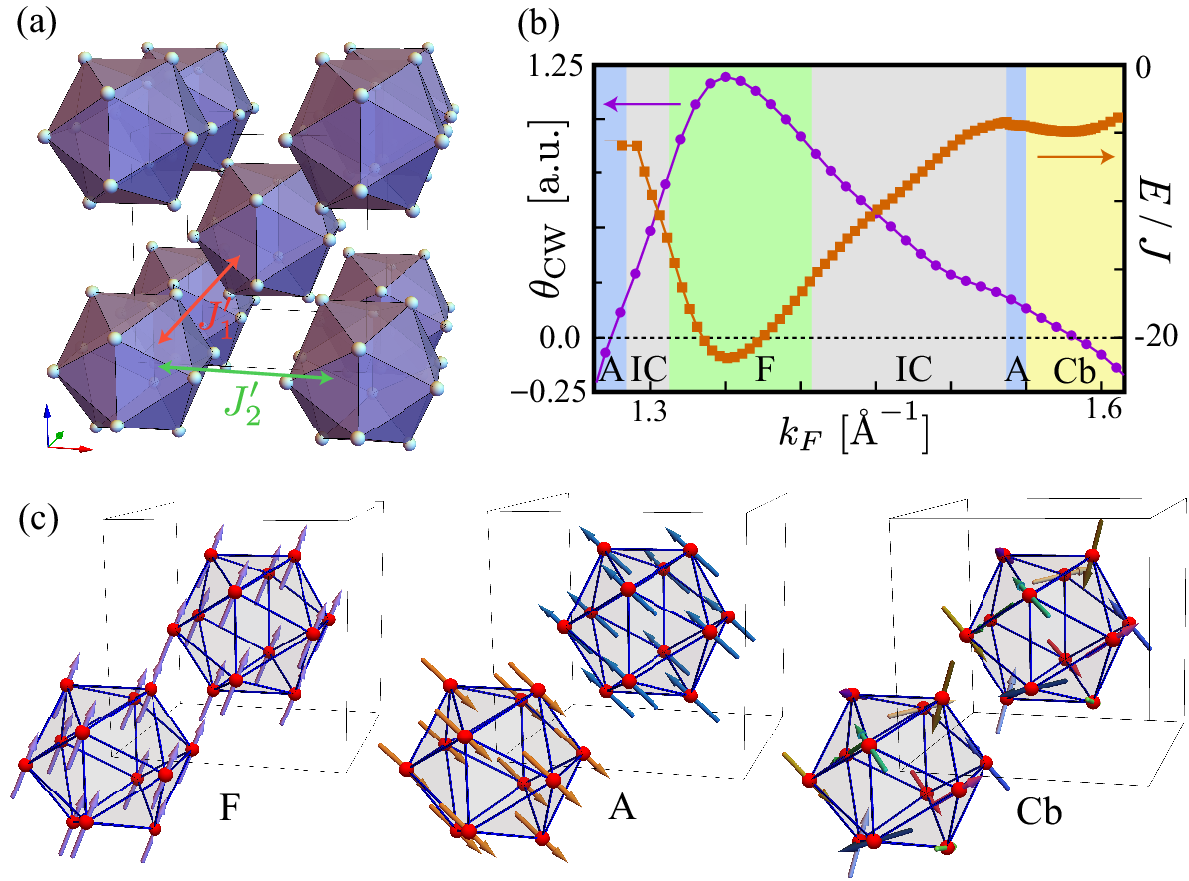}
\caption{(a) Crystal structure of Tsai-type 1/1 AC. Icosahedral clusters consist of bcc structure, where magnetic moments are located at the vertices of icosahedra. Red and green arrows denote the 1st and 2nd neighbor clusters based on the cluster at the origin. (b) Magnetic phase diagram with the Curie-Weiss temperature $\tht_{\mrm{CW}}$ and the ground-state energy $E/J$, determined by the CMC simulation in the AC model~\cite{Miyazaki2020}. (c) Magnetic structures in the F, A, and Cb phases. Different colors of arrows correspond to different directions.}
\label{fig1}
\end{figure}

Contrary to QCs, ACs retain translational symmetry, while a large unit cell in ACs has the same local structure as QCs.
Hence, the ACs are suitable for theoretical studies of magnetic properties in QCs.
For this sake, two of the present authors together with others have conducted classical Monte-Carlo (CMC) simulations for an AC model of a Gd-based Tsai-type 1/1 AC~\cite{Miyazaki2020}, where magnetic moments of Gd ions at the vertices of icosahedrons composed of the bcc structure [see \figref{fig1}(a)] are coupled each other via the so-called Ruderman--Kittel--Kasuya--Yosida (RKKY) interaction~\cite{Sugimoto2016}.
Since a Gd ion has zero orbital angular momentum for localized $f$ electrons, single-ion anisotropy can be ignored in the Gd system.
Therefore, the AC model for the Gd system contains magnetic interactions among localized magnetic moments $\bm{S}_{\bm{r}}$ defined as $\ham=-\sum_{|\bm{r}-\bm{r}^\pri|<R_c} J_{|\bm{r}-\bm{r}^\pri|}\, \bm{S}_{\bm{r}}\cdot\bm{S}_{\bm{r}^\pri}$ with the RKKY interaction $J_{R}=J f(2k_F R)$, the Friedel oscillation $f(x)=(-x\cos x+\sin x)/x^4$, the Fermi wavenumber $k_F$, and the cutoff length $R_c$.
The CMC calculations for a sufficiently large system have yielded a phase diagram and magnetic orders shown in \figref{fig1}(b) and \figref{fig1}(c), respectively~\cite{Miyazaki2020}, where antiferromagnetic (A), incommensurate (IC), ferromagnetic (F), and cuboc (Cb) phases appear.
The resulting phase diagram qualitatively explained an experimentally reported behavior~\cite{Suzuki2021-2}.

\begin{figure}[htb]
\centering
\includegraphics[width=0.50\textwidth]{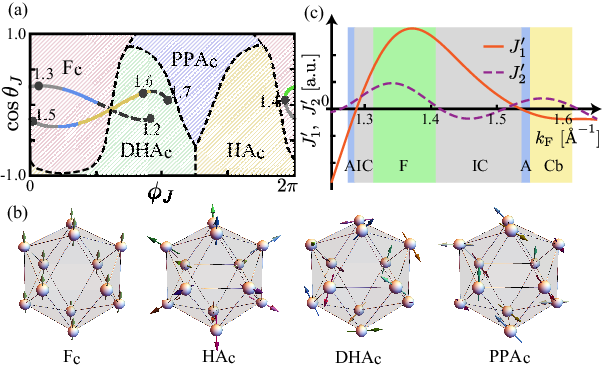}
\caption{(a) Phase diagram of magnetic orders in the cluster model.  $(J_1,J_2,J_3)=(\sin\tht_J\cos\phi_J,\sin\tht_J\sin\phi_J,\cos\tht_J)$ are used~\cite{Suzuki2021}. Red, yellow, green, and purple regions represent Fc, HAc, DHAc, and PPAc orders, respectively. Colored line denotes approximate parameter values for the Fermi wavenumber $k_F\in [1.2,1.7]\,\mrm{\AA}^{-1}$ according to the RKKY interaction. Colors on the line correspond to colors of phases in the AC model: blue (A), gray (IC), green (F), and yellow (Cb), respectively. (b) Magnetic structures in the cluster model. (c) Effective inter-cluster interactions, $J_1^\pri$ and $J_2^\pri$, for the 1st (orange solid line) and 2nd (purple dashed line) neighbor clusters, respectively [see \figref{fig1}(a)]. Colored regions correspond to phases in the AC model.}
\label{fig2}
\end{figure}

Even though a qualitative explanation of the experimental behavior was given by the AC model, full understanding of the mechanism of magnetic orders was not completed.
To deepen our understanding, three of the present authors have also examined a cluster model, where only 12 localized spins located at the vertices of an icosahedral cluster are considered and they are mutually coupled each other via the first (1st), second (2nd), and third (3rd) neighbor exchange interactions, $J_1$, $J_2$, and $J_3$, respectively, parametrized by two angles $\phi_J$ and $\theta_J$~\cite{Suzuki2021} (see the caption of \figref{fig2}).
In this model, only four phases of the cluster have been found: ferromagnetic (Fc), hedgehog antiferromagnetic (HAc), dual-hedgehog antiferromagnetic (DHAc), and parallel-pairs' antiferromagnetic (PPAc) orders [see \figref{fig2}(a) and \figref{fig2}(b)].
Note that definition of the signs of $J_1$, $J_2$, and $J_3$ in \figref{fig2} differs from that in the previous study~\cite{Suzuki2021}.

As mentioned above, the present authors investigated the magnetic orders in Gd-based ACs by using two models separately: the AC model and the cluster model.
However, relations between the two models have not been yet clarified.
In this short note, we show an intuitive understanding of emerging magnetic orders in the AC model by using the cluster model but with effective inter-cluster interactions.

In \figref{fig2}(a), possible phases in the cluster model are shown by changing the two angles $\phi_J$ and $\theta_J$~\cite{Suzuki2021}, where the values of $J_1$, $J_2$, and $J_3$ are determined from the angles.
In Gd-based ACs, $J_1$, $J_2$, and $J_3$ are determined by corresponding bond lengths in an icosahedron according to the RKKY interactions mentioned above.
We plot approximate values of $\phi_J$ and $\theta_J$ evaluated from $J_1$, $J_2$, and $J_3$ for Gd-based ACs with $k_F\in [1.2,1.7]\,\mrm{\AA}^{-1}$ on \figref{fig2}(a) as a colored line (note 2$\pi$ periodicity of $\phi_J$).
The color on the line corresponds to the color of each phase obtained using the AC model as shown in \figref{fig1}(b).
We find that a segment of the line for $k_F\in [1.25,1.55]\,\mrm{\AA}^{-1}$, which corresponds to all phases except for the Cb phase in \figref{fig1}(b), is located at the ferromagnetic Fc phase in the cluster model as shown in \figref{fig2}(a).
Therefore, the emergence of the A and IC phases in the AC model is attributed to the presence of inter-cluster interactions.
An effective inter-cluster interaction $J_1^\pri$ ($J_2^\pri$) for the Fc phase can be evaluated by summing over all possible RKKY interactions between the cluster at the origin and its 1st (2nd) neighbor cluster as shown by red (green) arrow in \figref{fig1}(a).
The resulting $J_1^\pri$ and $J_2^\pri$, denoted by solid and broken lines, respectively, in \figref{fig2}(c), show opposite signs in the IC phase and A phases in the AC model [see colored regions in \figref{fig2}(c)].

Combining the results in the AC and cluster models, we give below an intuitive explanation on the magnetic orders of the Gd-based Tsai-type 1/1 AC obtained by the AC model.\\
{\it The F phase.} The intra-cluster interactions favor the Fc order in a cluster.
Moreover, both $J_1^\pri$ and $J_2^\pri$ are ferromagnetic ($J_1^\pri>0$ and $J_2^\pri>0$) as shown in \figref{fig2}(c), resulting in trivial ferromagnetism in the AC.\\
{\it The A phases.} The Fc order is favored in a cluster, while the inter-cluster interactions are different from the F phase: $J_1^\pri<0$ and $J_2^\pri>0$.
Hence, there appears a two-sublattice antiferromagnetic structure consisting of the body-centered and cube-cornered clusters like the N\'{e}el structure in the bcc lattice.
Thus, we can call this {\it cluster N\'{e}el} order.\\
{\it The IC phases.} The intra-cluster interactions prefer the Fc order, whereas magnetic frustration exists.
In the IC phase for $k_F\in [1.41,1.54]\,\mrm{\AA}^{-1}$ in \figref{fig2}(c), $J_1^\pri>0$ and $J_2^\pri<0$.
Hence, magnetic frustration occurs in a triangle formed by the 1st-1st-2nd neighbor inter-cluster bonds.
On the contrary, in the IC phase for $k_F\in [1.28,1.32]\,\mrm{\AA}^{-1}$, there is magnetic frustration in a plaquette formed by the 1st neighbor inter-cluster bonds and the intra-cluster bonds.
In this IC phase, all the intra-cluster interactions are ferromagnetic, while there are comparable ferromagnetic and antiferromagnetic interactions in the 1st neighbor inter-cluster interactions as evidenced from zero crossing of $J_1^\pri$.
Thus, there is a plaquette consisting of two ferromagnetic intra-cluster bonds and comparable ferromagnetic and antiferromagnetic inter-cluster bonds in the 1st neighboring cluster, inducing magnetic frustration.\\
{\it The Cb phase.} This phase corresponds to the DHAc phase in the cluster model.
Since the net moment of the DHAc order is zero, effective inter-cluster interactions are not a good quantity to be discussed.
Rather inter-cluster RKKY interactions but only for short distance (without summation) seem to be crucial.
These interactions are antiferromagnetic for both the 1st and 2nd neighbor clusters.
In the DHAc order, magnetic moments on opposite side of cluster are anti-parallel.
Hence, the magnetic moments facing neighboring clusters can mediate the ferroic order between the two DHAc clusters.
Consequently, the Cb order obtained in the AC model is a consequence of the DHAc order in the cluster model, although high symmetry in the DHAc order is reduced to the lower one in the Cb order in the 1/1 AC.

In summary, we give an intuitive explanation of magnetic orders in the 1/1 ACs based on the cluster model including inter-cluster interactions.
We have found that the F, A, and IC phases in the AC model correspond to the Fc phase in the cluster model without inter-cluster interactions.
Therefore, the presence of the A and IC phases are due to the presence of effective inter-cluster interactions.
Introducing only the 1st and 2nd neighbor effective inter-cluster interactions gives us a consistent understanding of magnetic phases in the 1/1 ACs.
In contrast, the Cb phase in the AC model corresponds to the DHAc phase in the cluster model.

\begin{acknowledgment}
This work was supported by Challenging Research (Exploratory) (Grant No.~JP17K18764), Grant-in-Aid for Scientific Research on Innovative Areas (Grant No.~JP19H05821), and Grant-in-Aid for Scientific Research (B) (Grant No.~JP21H01044).
\end{acknowledgment}

\bibliographystyle{jpsj}
\bibliography{refs-a}


\end{document}